\documentclass{PoS}

\usepackage{amsmath,amssymb,bm,graphics}

\newcommand{\mbf}[1]{\mathbf{#1}}
\renewcommand{\bar}[1]{\overline{#1}}

\title{AdS/QCD, Light-Front Holography, and Color Confinement}

\ShortTitle{Light-Front Holography}

\author{\speaker{Stanley J. Brodsky}\thanks{This research was supported by the Department of Energy contract DE--AC02--76SF00515.}\\
        SLAC National Accelerator Laboratory\\
        Stanford University, Stanford, CA 94309, USA\\
        E-mail: \email{sjbth@slac.stanford.edu}}

\author{Guy F. de T\'eramond \\
      Universidad de Costa Rica, San Jos\'e, Costa Rica\\
       E-mail: \email{gdt@asterix.crnet.cr}}

\abstract{A  remarkable holographic feature of dynamics in AdS space in five dimensions is that it is dual to  Hamiltonian theory in physical space-time,  quantized at fixed light-front time
$\tau  = t+z/c$.   This light-front holographic principle provides a precise relation between the bound-state amplitudes in  
AdS space and the boost-invariant light-front wavefunctions describing the internal structure of hadrons in physical space-time.  The fifth dimension coordinate $z$ is dual to the light front variable $\zeta $  describing the 
invariant separation  of the quark constituents. The resulting valence Fock-state wavefunction eigensolutions
of the light-front QCD Hamiltonian satisfy a single-variable relativistic equation of motion, analogous to the nonrelativistic radial Schr\"odinger equation.  The soft-wall dilaton profile $e^{\kappa^2 z^2}$  provides  a model for the light-front potential which is color-confining and reproduces well the linear Regge behavior of the light-quark hadron spectrum in both $L$ the orbital angular momentum and $n$ the radial node number.   The pion mass vanishes in the chiral limit and other features of chiral symmetry are satisfied. The resulting running  QCD coupling displays an infrared fixed point.
The elastic and transition form factors of the pion and the nucleons are also found to be well described in this framework.  
The light-front AdS/QCD  holographic approach thus gives  a frame-independent analytic first approximation of the color-confining dynamics,  spectroscopy, and excitation spectra of relativistic light-quark bound states in QCD.    \
          }

\FullConference{Xth Quark Confinement and the Hadron Spectrum,\\
		October 8-12, 2012\\
		TUM Campus Garching, Munich, Germany}

\begin{document}

\section{Introduction}

The AdS/QCD approach to hadron physics, combined with light-front holography, provides a remarkable description of both hadron spectroscopy and dynamics.  It is relativistic, frame-independent, color-confining, and analytic. It also successfully predicts many phenomenological features of QCD, such as space-like and time-like form factors, scaling laws, Regge spectroscopy, and has many properties associated with chiral symmetry.  In this approach, baryons have three colors, and constraints from the operator product at short distances are satisfied. Light-front (LF) holography predicts not only the hadron spectrum, but also the analytic form of the light-front wave functions (LFWFs) of hadrons, thus allowing many new phenomenological predictions and tests. It also gives the form of the running QCD coupling in the nonperturbative domain and provides new insight into the fate of gluons at soft scales.
As we discuss below, the equations of motion in AdS space have a remarkable holographic mapping to the equations of motion obtained in light-front Hamiltonian theory in physical space-time. Thus, in principle, one can compute physical observables in a strongly coupled gauge theory  in terms of an effective classical gravity theory.

The quantization of QCD at fixed light-front time~\cite{Dirac:1949cp} (Dirac's Front Form)  provides a first-principles Hamiltonian method for solving nonperturbative QCD. It is rigorous, has no fermion-doubling, is formulated in Minkowski space, and it is frame-independent.  Given the light-front wavefunctions $\psi_{n/H}$, one can
compute a large range of hadron
observables, starting with structure functions, generalized parton distributions, and form factors.
It is also possible to compute jet hadronization at the amplitude level from first principles from the LFWFs.~\cite{Brodsky:2008tk} A similar method has been used to predict the production of antihydrogen from the off-shell coalescence of relativistic antiprotons and positrons.~\cite{Munger:1993kq}
The LFWFs of hadrons thus provide a direct connection between observables and the QCD Lagrangian. 

Solving nonperturbative QCD is equivalent to solving the light-front Heisenberg matrix eigenvalue problem.   Angular momentum $J^z$ is conserved at every vertex.
The LF vacuum is defined as the state of lowest invariant mass and is trivial up to zero modes.  There are thus no quark or gluon vacuum condensates in the LF vacuum-- the corresponding physics is contained within the 
LFWFs themselves,~\cite{Brodsky:2009zd,  Brodsky:2010xf} thus eliminating a major contribution to the cosmological constant.

The simplicity of the front form contrasts with the usual instant-form formalism. Current matrix elements defined at ordinary time $t$ must include the coupling of the photons and vector bosons fields  to connected vacuum currents; otherwise, the result is not Lorentz-invariant.  Thus the knowledge of the hadronic eigensolutions of the instant-form Hamiltonian are insufficient for determining form factors or other observables.   In addition, the boost of an instant form wavefunction from $p$ to $p+q$ changes particle number and is an extraordinarily complicated dynamical problem. 

The LF Hamiltonian method thus provides a first-principle method for solving nonperturbative QCD. It is rigorous, has no fermion-doubling, is formulated in Minkowski space, and it is frame-independent.  AdS/QCD provides a remarkable, analytic, first approximation to the LF approach which is systematically improvable using  a basis light-front quantization (BLFQ)~\cite{Vary:2009gt} and other methods.

It is remarkable fact that AdS/QCD,  which was originally motivated by
the AdS/CFT correspondence between gravity on a higher-dimensional anti--de Sitter (AdS) space and conformal field theories (CFT) in physical space-time,~\cite{Maldacena:1997re}  has a direct holographic mapping to light-front Hamiltonian theory.~\cite{deTeramond:2008ht}  
For example, the equation of motion for mesons on the light-front has exactly the same single-variable form as the AdS/QCD equation of motion; one can then interpret the AdS fifth dimension variable $z$ in terms of the physical variable $\zeta$, representing the invariant separation of the $q$ and $\bar q$ at fixed light-front time.  The AdS mass  parameter $\mu R$ maps to the LF orbital angular momentum.  The formulae for electromagnetic~\cite{Polchinski:2002jw} and gravitational~\cite{Abidin:2008ku} form factors in AdS space map to the exact Drell-Yan-West formulae in light-front QCD.~\cite{Brodsky:2006uqa, Brodsky:2007hb, Brodsky:2008pf}  Thus, as we shall show, the light-front holographic approach provides an analytic frame-independent first approximation to the color-confining dynamics,  spectroscopy, and excitation spectra of the relativistic light-quark bound states of QCD.

\section{The Light-Front Schr\"odinger Equation: A Semiclassical Approximation to QCD \label{LFQCD}}

Unlike ordinary instant-time quantization, light-front Hamiltonian equations of motion are frame independent; remarkably, they  have a structure which matches exactly the eigenmode equations in AdS space. This makes a direct connection of QCD with AdS methods possible.  In fact, one can
derive the light-front holographic duality of AdS  by starting from the light-front Hamiltonian equations of motion for a relativistic bound-state system
in physical space-time.~\cite{deTeramond:2008ht}
To a first semiclassical approximation, where quantum loops and quark masses
are not included, this leads to a LF Hamiltonian equation which
describes the bound state dynamics of light hadrons  in terms of
an invariant impact variable $\zeta$, which measures the
separation of the partons within the hadron at  fixed light-front time $\tau = t+z/c$.~\cite{Dirac:1949cp} 
This allows one to identify the  variable $z$ in
AdS space with the impact variable $\zeta$,~\cite{Brodsky:2006uqa,
Brodsky:2008pf, deTeramond:2008ht}  thus giving  the holographic
variable a precise definition and very intuitive meaning in light-front QCD.  In the case of a two-parton state, $\zeta^2= x(1-x)\mbf{b}_\perp^2$.

It is advantageous to reduce the full multiparticle eigenvalue problem of the LF Hamiltonian to an effective light-front Schr\"odinger equation (LFSE) which acts on the valence sector LF wavefunction and determines each eigensolution separately.~\cite{Pauli:1998tf}   In contrast,  diagonalizing the LF Hamiltonian yields all eigensolutions simultaneously, a complex task.
The central problem for deriving the LFSE becomes the derivation of the effective interaction $U$ which acts only on the valence sector of the theory and has, by definition, the same eigenvalue spectrum as the initial Hamiltonian problem. For carrying out this program one must systematically express the higher Fock components as functionals of the lower ones. The method has the advantage that the Fock space is not truncated and the symmetries of the Lagrangian are preserved.~\cite{Pauli:1998tf}
One starts with the QCD Lagrangian and derives the light-front Hamiltonian.~\cite{Brodsky:1997de}
One can then transform the LF equation of motion to a semiclassical QCD equation operating on the lowest Fock state of the hadron. There are no approximations at this point, and the result is relativistic and frame-independent.

For example, we can write the $q \bar q$ LFSE for a meson  as
$\psi(x,\zeta, \varphi) = e^{i L \varphi} X(x) \frac{\phi(\zeta)}{\sqrt{2 \pi \zeta}},$
thus factoring the angular dependence $\varphi$ and the longitudinal, $X(x)$, and transverse mode $\phi(\zeta)$.
In the limit of zero quark masses the longitudinal mode decouples and
the LF eigenvalue equation $P_\mu P^\mu \vert \phi \rangle  =  M^2 \vert \phi \rangle$
is thus a light-front  wave equation for $\phi$~\cite{deTeramond:2008ht}
\begin{equation} \label{LFWE}
\left[-\frac{d^2}{d\zeta^2}
- \frac{1 - 4L^2}{4\zeta^2} + U\left(\zeta^2, J, M^2\right) \right]
\phi_{J,L,n}(\zeta^2) = M^2 \phi_{J,L,n}(\zeta^2),
\end{equation}
a relativistic {\it single-variable}  LF  Schr\"odinger equation.

The potential in the LFSE is determined from the two-particle irreducible (2PI) $ q \bar q \to q \bar q $ Greens' function.  In particular, the higher Fock states in intermediate states
leads to an effective interaction $U(\zeta^2, J ,M^2)$  for the valence $\vert q \bar q \rangle$ Fock state.~\cite{Pauli:1998tf}
The potential $U$ thus depends on the hadronic eigenvalue $M^2$ via the LF energy denominators 
$P^-_{\rm initial} - P^-_{\rm intermediate} + i \epsilon$
of the intermediate states which connect different LF Fock states.  
Here $P^-_{\rm initial} =( {M^2 + \mathbf{P}^2_\perp)/ P^+}$. The dependence of $U$ on $M^2$ is analogous to the retardation effect in QED interactions, such as the hyperfine splitting in muonium, which involves the exchange of a propagating photon.  Accordingly, the eigenvalues $M^2$ must be determined 
self-consistently.~\cite{deTeramond:2012cs} 
The  $M^2$ dependence of the effective potential thus reflects the contributions from higher Fock states in the LFSE (\ref{LFWE}),  since
 $U(\zeta^2, J ,M^2)$ is also the kernel for the  scattering amplitude $q \bar q \to q \bar q$ at $s = M^2.$    It has only ``proper'' contributions; {\it i.e.}, it has no $q \bar q$ intermediate state.  The potential can be constructed, in principle systematically, using LF time-ordered perturbation theory.  In fact, as we shall see below, the QCD theory has the identical form as the AdS theory, but with the quantum field-theoretic corrections due to the higher Fock states giving a general form for the potential.  This provides a novel way to solve nonperturbative QCD.  The LFSE for QCD becomes increasingly accurate as one includes contributions from very high particle number Fock states.  In the case of QED, this provides  a new approach to the spectroscopy of positronium and other atomic bound states.

The above discussion assumes massless quarks. More generally we must include quark mass terms,~\cite{Brodsky:2008pg,Gutsche:2011uj}
${m^2_a / x}  + {m^2_b/(1-x)}$, in the kinetic energy term and allow  the potential $  U(\zeta^2, x, J,M^2)$ to have dependence on the LF momentum fraction $x$.  The quark masses also appear in $U$ due to the  presence in the LF denominators as well as the chirality-violating interactions connecting the valence Fock state to the higher Fock states. In this case, however, the equation of motion cannot be reduced to a single variable.

The LFSE approach also can be applied to atomic bound states in QED and nuclei. In principle, one could compute the spectrum and dynamics of atoms, such as the Lamb shift and hyperfine splitting of hydrogenic atoms to high precision by a systematic treatment of the potential. Unlike the ordinary  instant form, the resulting LFWFs are independent of the total momentum and can thus describe ``flying atoms'' without the need for dynamical boosts, 
such as the ``true muonium'' $(\mu^+ \mu^-)$ bound states which can be produced  by Bethe-Heitler pair production $\gamma \, Z \to (\mu^+ \mu^-)  \, Z$  below 
threshold.~\cite{Brodsky:2009gx} A related approach for determining the valence light-front wavefunction and studying the effects of higher Fock states without truncation has been given in Ref.~\cite{Chabysheva:2011ed}.

\section{Effective Confinement Interaction from the Gauge/gravity Correspondence}

Recently we have derived holographic wave equations for hadrons with arbitrary spin starting from an effective action in a higher-dimensional space asymptotic to  AdS space.~\cite{deTeramond:2013it}    An essential element is the mapping of the higher-dimensional equations of motion to the light-front Hamiltonian equation for relativistic bound-states  found in Ref.~\cite {deTeramond:2008ht}.  This procedure allows a clear distinction between the kinematical and dynamical aspects of the light-front holographic approach to hadron physics.  Accordingly, the non-trivial geometry of pure AdS space encodes the kinematics,  and the additional deformations of AdS space encode the dynamics, including confinement.~\cite{deTeramond:2013it}

A spin-$J$ field in AdS$_{d+1}$ is represented by a rank $J$ tensor field $\Phi_{M_1 \cdots M_J}$, which is totally symmetric in all its indices.  In presence of a dilaton background field $\varphi(z)$ the effective action is~\cite{deTeramond:2013it} 
\begin{multline}
\label{Seff}
S_{\it eff} = \int d^{d} x \,dz \,\sqrt{\vert g \vert}  \; e^{\varphi(z)} \,g^{N_1 N_1'} \cdots  g^{N_J N_J'}   \Big(  g^{M M'} D_M \Phi^*_{N_1 \dots N_J}\, D_{M'} \Phi_{N_1 ' \dots N_J'}  \\
 - \mu_{\it eff}^2(z)  \, \Phi^*_{N_1 \dots N_J} \, \Phi_{N_1 ' \dots N_J'} \Big),
 \end{multline}
where the indices $M, N = 0, \cdots , d$, $\sqrt{g} = (R/z)^{d+1}$ and $D_M$ is the covariant derivative which includes parallel transport. The coordinates of AdS are the Minkowski coordinates $x^\mu$ and the holographic variable $z$, $x^M = \left(x^\mu, z\right)$. 
The effective mass  $\mu_{\it eff}(z)$, which encodes kinematical aspects of the problem, is an {\it a priori} unknown function,  but the additional symmetry breaking due to its $z$-dependence allows a clear separation of kinematical and dynamical effects.~\cite{deTeramond:2013it}  
The dilaton background field $\varphi(z)$ in  (\ref{Seff})   introduces an energy scale in the five-dimensional AdS action, thus breaking conformal invariance. It  vanishes in the conformal ultraviolet limit $z \to 0$.

 A physical hadron has plane-wave solutions and polarization indices along the 3 + 1 physical coordinates
 $\Phi_P(x,z)_{\nu_1 \cdots \nu_J} = e^{ i P \cdot x} \Phi_J(z) \epsilon_{\nu_1 \cdots \nu_J}({P})$,
 with four-momentum $P_\mu$ and  invariant hadronic mass  $P_\mu P^\mu \! = M^2$. All other components vanish identically. 
 The wave equations for hadronic modes follow from the Euler-Lagrange equation for tensors orthogonal to the holographic coordinate $z$,  $\Phi_{z N_2 \cdots N_J}  = 0$. Terms in the action which are linear in tensor fields, with one or more indices along the holographic direction, $\Phi_{z N_2 \cdots N_J}$, give us 
 the kinematical constraints required to eliminate the lower-spin states.~\cite{deTeramond:2013it}  Upon variation with respect to $ \hat \Phi^*_{\nu_1 \dots \nu_J}$,
 we find the equation of motion~\cite{deTeramond:2013it}  
\begin{equation}  \label{PhiJM}
 \left[ 
   -  \frac{ z^{d-1- 2J}}{e^{\varphi(z)}}   \partial_z \left(\frac{e^{\varphi(z)}}{z^{d-1-2J}} \partial_z   \right) 
  +  \frac{(m\,R )^2}{z^2}  \right]  \Phi_J = M^2 \Phi_J,
  \end{equation}
  with  $(m \, R)^2 =(\mu_{\it eff}(z) R)^2  - J z \, \varphi'(z) + J(d - J +1)$,
  which is  the result found in Refs.~\cite{deTeramond:2008ht, deTeramond:2012rt} by rescaling the wave equation for a scalar field. Similar results were found in
  in Ref.~\cite{Gutsche:2011vb}.
 Upon variation with respect to
$ \hat \Phi^*_{N_1 \cdots z  \cdots N_J}$  we find the kinematical constraints which  eliminate lower spin states from the symmetric field tensor~\cite{deTeramond:2013it}  
\begin{equation} \label{sub-spin}
 \eta^{\mu \nu } P_\mu \,\epsilon_{\nu \nu_2 \cdots \nu_J}({P})=0, \quad
\eta^{\mu \nu } \,\epsilon_{\mu \nu \nu_3  \cdots \nu_J}({P})=0.
 \end{equation}

Upon the substitution of the holographic variable $z$ by the LF invariant variable $\zeta$ and replacing
  $\Phi_J(z)   = \left(R/z\right)^{J- (d-1)/2} e^{-\varphi(z)/2} \, \phi_J(z)$ 
in (\ref{PhiJM}), we find for $d=4$ the LFSE (\ref{LFWE}) with effective potential~\cite{deTeramond:2010ge}
\begin{equation} \label{U}
U(\zeta^2, J) = \frac{1}{2}\varphi''(\zeta^2) +\frac{1}{4} \varphi'(\zeta^2)^2  + \frac{2J - 3}{2 \zeta} \varphi'(\zeta^2) ,
\end{equation}
provided that the AdS mass $m$ in (\ref{PhiJM}) is related to the internal orbital angular momentum $L = max \vert L^z \vert$ and the total angular momentum $J^z = L^z + S^z$ according to $(m \, R)^2 = - (2-J)^2 + L^2$.  The critical value  $L=0$  corresponds to the lowest possible stable solution, the ground state of the LF Hamiltonian.
For $J = 0$ the five dimensional mass $m$
 is related to the orbital  momentum of the hadronic bound state by
 $(m \, R)^2 = - 4 + L^2$ and thus  $(m\, R)^2 \ge - 4$. The quantum mechanical stability condition $L^2 \ge 0$ is thus equivalent to the Breitenlohner-Freedman stability bound in AdS.~\cite{Breitenlohner:1982jf}

The correspondence between the LF and AdS equations  thus determines the effective confining interaction $U$ in terms of the infrared behavior of AdS space and gives the holographic variable $z$ a kinematical interpretation. The identification of the orbital angular momentum 
is also a key element of our description of the internal structure of hadrons using holographic principles.

\begin{figure}[h]
\centering
\includegraphics[width=6.15cm]{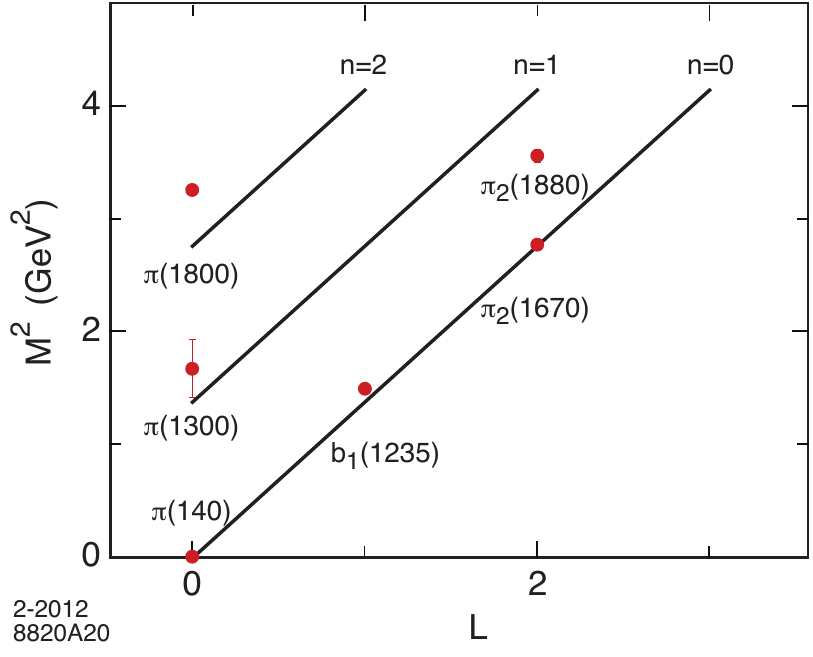}  \hspace{0pt}
\includegraphics[width=6.15cm]{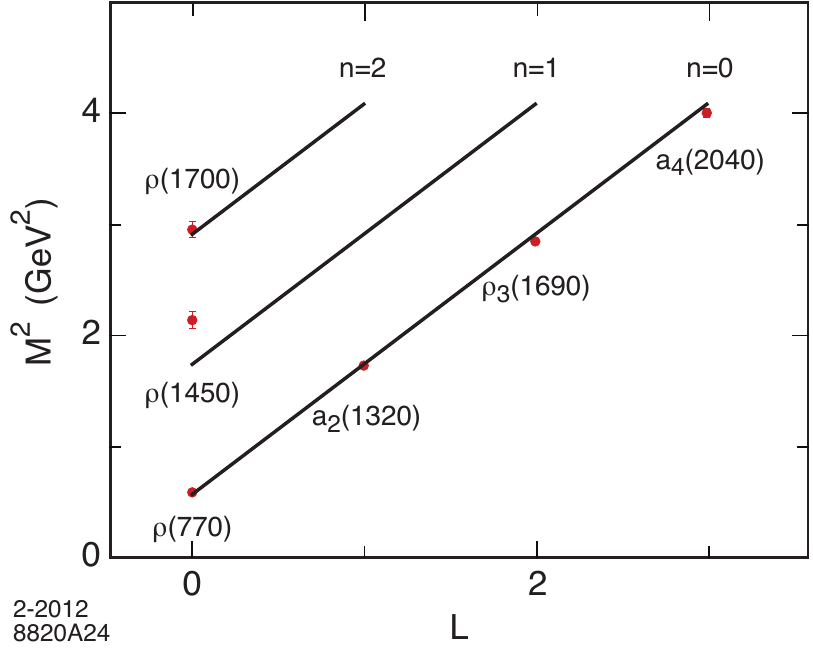}
 \caption{$I = 1$ parent and daughter Regge trajectories for the $\pi$-meson family (left) with
$\kappa= 0.59$ GeV; and  the   $\rho$-meson
 family (right) with $\kappa= 0.54$ GeV.}
\label{pionspec}
\end{figure} 

\section{A Soft-Wall Holographic Model}

The soft-wall model  $\varphi(z) = \pm \kappa^2 z^2$
leads to linear Regge trajectories~\cite{Karch:2006pv} and avoids the ambiguities in the choice of boundary conditions at the infrared wall.   However, the solution $\varphi(z) =  -\kappa^2 z^2$ is incompatible with the light-front  constituent interpretation of hadronic states as shown in Ref.~\cite{deTeramond:2013it} . 
In fact, it can be shown that if one starts with a dilaton of the general form $\varphi(z, s) = \kappa z^s$, for arbitrary values of  $s$, the constraints imposed by chiral symmetry in the limit of massless quarks determine uniquely the value  $s = 2$.~\cite{Brodsky:2013yy}  This is a remarkable result, since this value corresponds precisely to the dilaton profile required to  reproduce the  linear Regge behavior.

\begin{figure}[h]
\begin{center}
 \includegraphics[width=6.15cm]{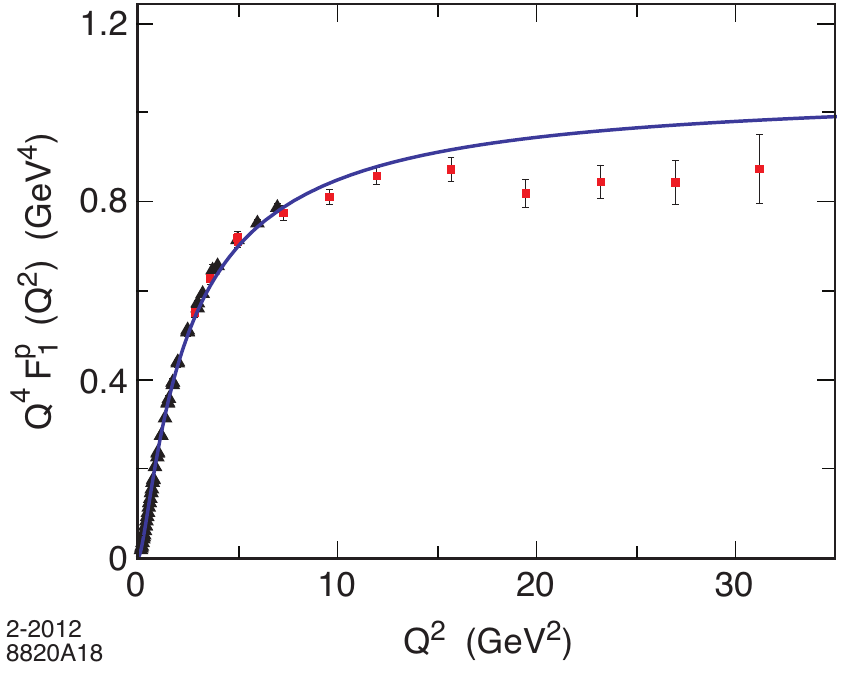}   \hspace{0pt}
\includegraphics[width=6.15cm]{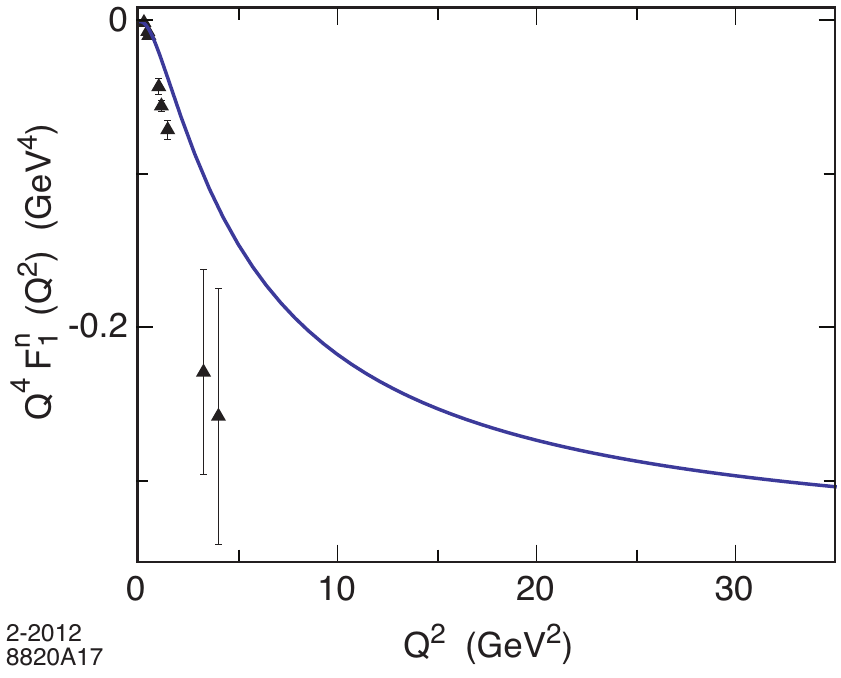}
 \includegraphics[width=6.15cm]{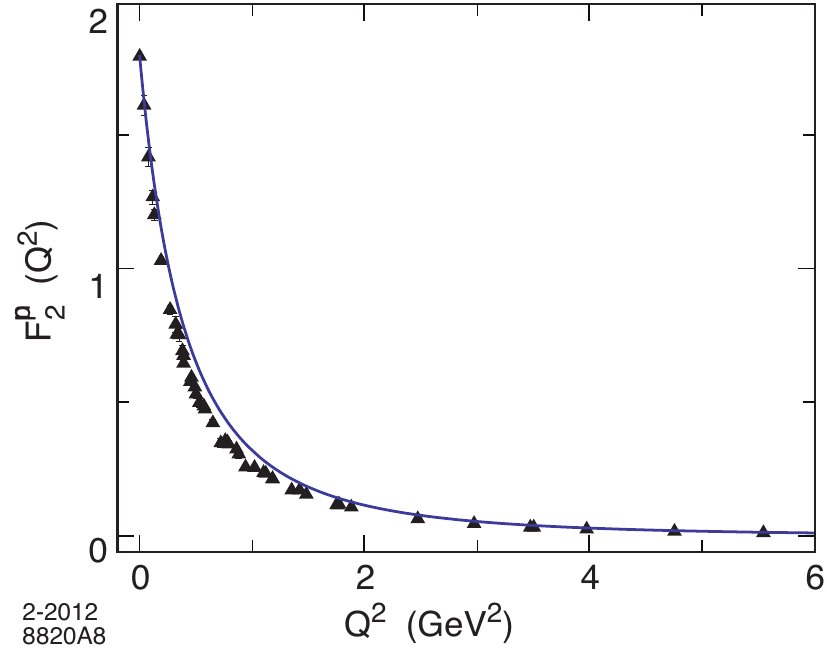}   \hspace{0pt}
\includegraphics[width=6.15cm]{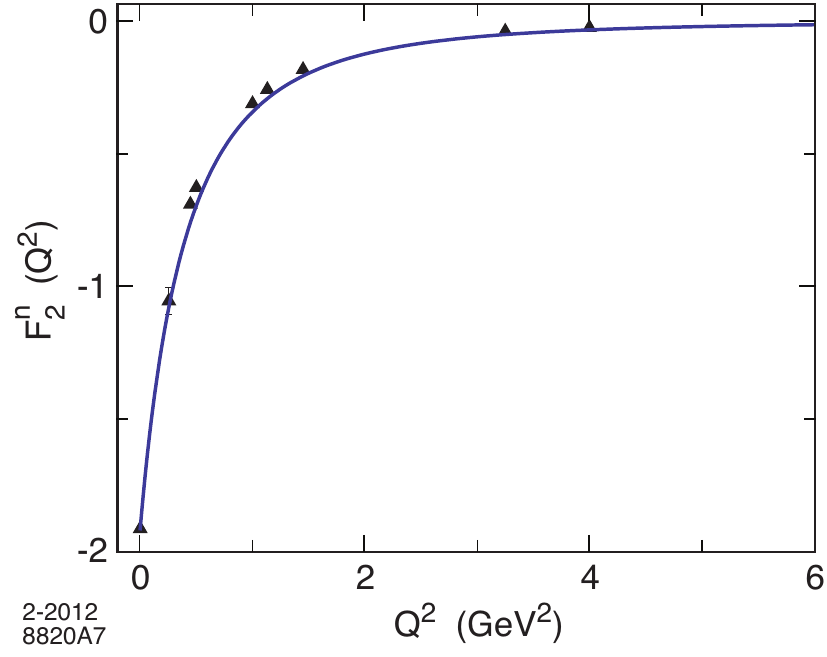}
 \caption{Predictions for  $Q^4 F_1^p(Q^2)$ (upper left) and   $Q^4 F_1^n(Q^2)$ (upper right) in the
soft wall model. The results for  $F_2^p(Q^2)$ (lower left)  and  $F_2^n(Q^2)$ (lower right) are normalized to static quantities. 
The nucleon spin-flavor structure corresponds to the $SU(6)$ limit.}
\label{fig:nucleonFF}
\end{center}
\end{figure}

The  confining solution $\varphi = \exp{\left(\kappa^2 z^2\right)}$ leads to the  effective potential 
$U(\zeta^2,J) =   \kappa^4 \zeta^2 + 2 \kappa^2(J - 1)$ and  Eq.  (\ref{LFWE}) has eigenvalues
$M_{n, J, L}^2 = 4 \kappa^2 \left(n + \frac{J+L}{2} \right)$,
with a string Regge form $M^2 \sim n + L$.  
A discussion of the light meson and baryon spectrum,  as well as  the elastic and transition form factors of the light hadrons using LF holographic methods, is given in 
Ref.~\cite{deTeramond:2012rt}.  As an example, the spectral predictions  for the $J = L + S$ light pseudoscalar and vector meson  states are  compared with experimental data in Fig. \ref{pionspec} for the positive sign dilaton model.  The data is from PDG.~\cite{PDG2012}   The results for $Q^4 F_1^p(Q^2)$ and $Q^4 F_1^n(Q^2)$  in the valence approximation are shown in Fig. \ref{fig:nucleonFF}.  We also show in  Fig. \ref{fig:nucleonFF}
the results for $F_2^p(Q^2)$ and $F_2^n(Q^2)$ for the same value of $\kappa$ normalized  to the static quantities $\chi_p$ and $\chi_n$.
To compare with physical data we have shifted the poles 
in the form factor to their physical values located at $M^2 = 4 \kappa^2(n + 1/2)$ 
following the  discussion in Ref.~\cite{deTeramond:2012rt}.
The value $\kappa = 0.545$ GeV  is determined from the $\rho$ mass.  The data compilation  is from Ref.~\cite{Diehl:2005wq}.

Holographic QCD also provide powerful nonperturbative analytical tools to extend the multi-component light-front Fock structure of amplitudes to the time-like region.
Consider for example the elastic pion form factor $F_\pi$. The pion is a superposition of an infinite number of Fock components  $\vert N \rangle$, $\vert \pi\rangle = \sum_N \psi_N \vert N \rangle$, and thus 
the pion form factor is  given by~\cite{deTeramond:2012rt}
$
F_\pi(s) = \sum_\tau P_\tau F_\tau(s),
$
where $P_\tau$ is the probability for  twist $\tau$ (the number of components $N$ in a given Fock-state) and  $F_\tau$
\begin{equation}   \label{Ftau} 
F_\tau(s) =  \frac{M^2_{n=0} M^2_{n=1} \cdots M^2_{n=\tau-2} }{{\left(M^2_{n=0} - s \right) }
 \left(M^2_{n=1} - s\right)  \cdots 
       \left(M^2_{n=\tau-2}  - s\right)},
\end{equation}   
is  written as a $\tau - 1$ product of poles.~\cite{Brodsky:2007hb}  Normalization at $Q^2 =0$,  $F_\pi(0) = 1$, implies that 
 $\sum_\tau P_\tau = 1$ if all possible states are included.
 
 In the strongly coupled semiclassical gauge/gravity limit, hadrons have zero widths and are stable,  as  in the  $N_C \to \infty$ limit of QCD.  One can nonetheless modify  (\ref{Ftau})  by introducing  finite widths  using the substitution 
 $s \to s + i \sqrt{s} \, \Gamma_n$. The resulting expression  has a series of resonance poles  in the non-physical lower-half complex $ \sqrt{s}$-plane (Im $\sqrt{s} <0$) located at  $\alpha^n_\pm = \frac{1}{2}\left(- i \Gamma_n \pm \sqrt{4 M_n^2 - \Gamma_n^2 } \right)$ with $n = 0, 1, \cdots, \tau-2$. In the upper half plane  (Im $\sqrt{s} >0$) --which maps to the physical sheet, the pion form factor is an analytic function and thus can be mapped to the space-like region by a $\pi/2$ rotation in the complex $\sqrt{s}$-plane such as to avoid cutting the real axis. This simple procedure allows us to analytically continue the holographic results to the space-like region where the pion form factor is also modified   by the finite widths of the vector meson resonances from interactions with the continuum.~\cite{Arriola:2012vk}

\begin{figure*}[h]
\centering 
\includegraphics[width=12cm]{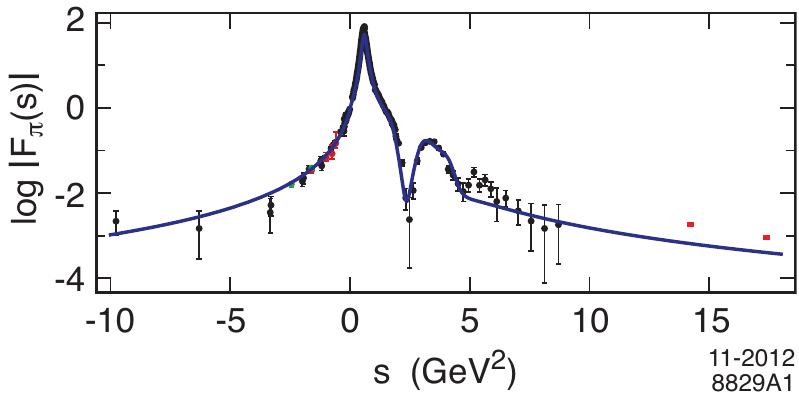}  \hspace{0pt}
\caption{Structure of the space-like ($s<0$)  and time-like ($s> 4 m_\pi^2$)  pion  form factor in light-front holography for a truncation of the pion wave function up to twist five.
The space-like data are  the compilation  from Baldini  {\it et al.}~\cite{Baldini:1998qn} (black)  and JLAB data~\cite{Tadevosyan:2007yd} (green and red). The time-like data
are from the recent precise measurements from BABAR~\cite{Aubert:2009ad} (black)  and CLEO~\cite{Seth:2012nn} (red).}
\label{pionFF}
\end{figure*} 

To illustrate the relevance of higher Fock states in the analytic  structure of the pion form factor we consider a simple phenomenological model where we include the first three components in a Fock expansion of the pion state $\vert \pi \rangle  = \psi^{L=0}_{q \bar q /\pi} \vert q \bar q; L=0   \rangle_{\tau=2} 
+  \psi^{L=0}_{q \bar q q \bar q} \vert q \bar q  q \bar q;  L=0  \rangle_{\tau=4}  +  \psi^{L=1}_{q \bar q q \bar q} \vert q \bar q  q \bar q; L =1\rangle_{\tau=5} + \cdots,$ and no constituent dynamical gluons.~\cite{Brodsky:2011pw}  The $J^{PC} = 0^{- +}$ twist-2, twist-4 and twist-5 states  are created by the interpolating operators $\mathcal{O}_2 = \bar q \gamma^+ \gamma_5  q$,  $\mathcal{O}_4 = \bar q \gamma^+ \gamma_5  q  \bar q q$ and $\mathcal{O}_5 = \bar q \gamma^+ \gamma_5  D^\mu q  \bar q \gamma_\mu q$. The results for the space-like and time-like form factor are shown in Fig. \ref{pionFF}. We choose the values $\Gamma_\rho =  149$ MeV,   $\Gamma_{\rho'} =  300$ MeV,  $\Gamma_{\rho''} =  250$ MeV and $\Gamma_{\rho'''} = 200$ MeV. The chosen width of the $\rho'$ is slightly smaller than the PDG value listed in Ref. \cite{PDG2012}. The widths of the $\rho$ and $\rho''$ are the PDG  values and the width of the $\rho'''$ is unknown. 
 The results correspond to $P^{L=0}_{q \bar q q \bar q}$ = 2.5 \% and  $P^{L=1}_{q \bar q q \bar q}$ = 4.5 \% the admixture of the
$\vert q \bar q q \bar q; L=0  \rangle$ and $\vert q \bar q q \bar q; L=1  \rangle$ states. The value of $P^{L=0}_{q \bar q q \bar q}$ and $P^{L=1} _{q \bar q q \bar q}$ (and the widths) are input in the model. 
The main features of the space-like and time-like regions are well described by the same formula with a minimal number of parameters. In contrast, models  based on hadronic degrees of freedom involve sums over a large number of intermediate states and thus require, for large $s$, a very large number of hadronic parameters~\cite{Kuhn:1990ad, Bruch:2004py, Hanhart:2012wi}.  A detailed data comparison requires the introduction  of thresholds and $s$-dependent widths from multiparticle states coupled to the $\rho$-resonances. This will be described elsewhere.~\cite{dTB}

\section{Conclusions}

Despite some limitations of AdS/QCD, the LF holographic  approach to the gauge/gravity duality, has  given significant physical insight into the strongly-coupled nature and internal structure of hadrons.  In particular, the light-front AdS/QCD soft-wall model provides an elegant analytic framework for describing nonperturbative  hadron dynamics, the systematics of the excitation spectrum of hadrons, including their empirical multiplicities and degeneracies. The pion is massless in the chiral limit and the light-hadron spectrum displays linear Regge trajectories with the same slope in the node number $n$ and orbital angular momentum $L$.   It also provides powerful new analytical tools for computing hadronic transition amplitudes incorporating conformal scaling behavior at short distances and the transition from the hard-scattering perturbative domain, where quark and gluons are the relevant degrees of freedom, to the long-range confining hadronic region.

The confining light-front potential for hadronic  bound-state equations for particles with arbitrary spin can be derived from an effective invariant action in the higher dimensional classical gravitational theory. The fact that we can map the equations of motion for arbitrary spin from the gravitational theory to a Hamiltonian equation of motion in light-front quantized QCD is a remarkable result. The undisturbed AdS geometry reproduces the kinematical aspects of the light-front Hamiltonian, notably the emergence of a LF angular momentum which is holographically identified with the mass in  the gravitational theory.    The breaking of the maximal symmetry of AdS  then allows the introduction of  the confinement dynamics  of the theory in physical space-time.
In order to fully  preserve all the kinematical aspects, a consistent mapping to LF quantized QCD requires a clear separation between the kinematical and dynamical effects.~\cite{deTeramond:2013it}

The effective interaction $ U(\zeta^2) = \kappa^4 \zeta^2 + 2 \kappa^2(J-1)$  that is derived from the AdS/QCD model is instantaneous in LF time and acts on the lowest state of the LF Hamiltonian.  This equation describes the spectrum of mesons as a function of $n$, the number of nodes in $\zeta^2$ and the total angular momentum $J=J^z$,
with $J^z = L^z + S^z$  the sum of the internal orbital angular momentum of the constituents. 
It is the relativistic, frame-independent front-form analog of the non-relativistic radial Schr\"odinger equation for muonium  and other hydrogenic atoms in the presence of an instantaneous Coulomb potential.
The 
holographic harmonic oscillator potential could in fact emerge from the exact QCD formulation when one includes contributions from the LFSE potential $U$ which are due to the exchange of two connected gluons; {\it i.e.}, ``H'' diagrams.~\cite{Appelquist:1977tw}
We notice that $U$ becomes complex for an excited state since a denominator can vanish; this gives a complex eigenvalue and the decay width.

\section*{Acknowledgments}

Section 3 is based on a collaboration with Hans Guenter Dosch, and we thank him for helpful comments.
This research was supported by the Department of Energy contract DE--AC02--76SF00515.
SLAC-PUB-15350
.


\begin{thebibliography}{99}

 \bibitem{Dirac:1949cp}
  P.~A.~M.~Dirac,
Rev.\ Mod.\ Phys.\  {\bf 21}, 392 (1949).


\bibitem{Brodsky:2008tk}
  S.~J.~Brodsky, G.~de Teramond and R.~Shrock,
  AIP Conf.\ Proc.\  {\bf 1056}, 3 (2008)
  [arXiv:0807.2484 [hep-ph]].


\bibitem{Munger:1993kq}
  C.~T.~Munger, S.~J.~Brodsky and I.~Schmidt,
  Phys.\ Rev.\  D {\bf 49}, 3228 (1994).
  
  
   \bibitem{Brodsky:2009zd}
  S.~J.~Brodsky and R.~Shrock,
  Proc.\ Nat.\ Acad.\ Sci.\  {\bf 108}, 45 (2011)
  [arXiv:0905.1151 [hep-th]].
  
  
\bibitem{Brodsky:2010xf}
  S.~J.~Brodsky, C.~D.~Roberts, R.~Shrock , P.~Tandy,
  Phys.\ Rev.\  {\bf C82}, 022201 (2010)
  [arXiv:1005.4610 [nucl-th]].
  
  
  \bibitem{Vary:2009gt} 
 J.~P.~Vary, H.~Honkanen, J.~Li, P.~Maris, S.~J.~Brodsky, A.~Harindranath, G.~F.~de Teramond and P.~Sternberg {\it et al.},
 Phys.\ Rev.\ C {\bf 81}, 035205 (2010)  [arXiv:0905.1411 [nucl-th]]. 


 \bibitem{Maldacena:1997re}
 J.~M.~Maldacena,
  Adv.\ Theor.\ Math.\ Phys.\  {\bf 2}, 231 (1998)
[{arXiv:hep-th/9711200}].


\bibitem{deTeramond:2008ht}
  G.~F.~de Teramond and S.~J.~Brodsky,
  Phys.\ Rev.\ Lett.\  {\bf 102}, 081601 (2009)
  [arXiv:0809.4899 [hep-ph]].
  
  
   \bibitem{Polchinski:2002jw}
  J.~Polchinski and M.~J.~Strassler,
 JHEP {\bf 0305}, 012 (2003)
  [{arXiv:hep-th/0209211}].
  

  \bibitem{Abidin:2008ku}
  Z.~Abidin and C.~E.~Carlson,
  {Phys.\ Rev.\  D {\bf 77}, 095007 (2008)}
  [{arXiv:0801.3839 [hep-ph]}].
  
  
   \bibitem{Brodsky:2006uqa}
  S.~J.~Brodsky and G.~F.~de Teramond,
  Phys.\ Rev.\ Lett.\  {\bf 96}, 201601 (2006)
  [{arXiv:hep-ph/0602252}];


 \bibitem{Brodsky:2007hb}
   S.~J.~Brodsky and G.~F.~de Teramond,
  Phys.\ Rev.\  D {\bf 77}, 056007 (2008)
 [arXiv:0707.3859 [hep-ph]].
 
 
 \bibitem{Brodsky:2008pf}
 S.~J.~Brodsky and G.~F.~de Teramond,
Phys.\ Rev.\  D {\bf 78}, 025032 (2008)


\bibitem{Pauli:1998tf}
  H.~C.~Pauli,
  Eur.\ Phys.\ J.\  C {\bf 7}, 289 (1999)
  [arXiv:hep-th/9809005].
  
  
  \bibitem{Brodsky:1997de}
  S.~J.~Brodsky, H.~C.~Pauli and S.~S.~Pinsky,
  Phys.\ Rept.\  {\bf 301}, 299 (1998)
  [arXiv:hep-ph/9705477].
  
  
  \bibitem{deTeramond:2012cs}
  G.~F.~de Teramond and S.~J.~Brodsky,
  arXiv:1206.4365 [hep-ph].
  
  
  \bibitem{Brodsky:2008pg}
  S.~J.~Brodsky and G.~F.~de Teramond,
  {World Scientific  Subnuclear Series, {\bf 45}, 139 (2007)}
  [{arXiv:0802.0514 [hep-ph]}].

  
  \bibitem{Gutsche:2011uj}
  T.~Gutsche, V.~E.~Lyubovitskij, I.~Schmidt and A.~Vega,
  Prog.\ Part.\ Nucl.\ Phys.\  {\bf 67}, 206 (2012)
  [arXiv:1111.5169 [hep-ph]].
  
  
  \bibitem{Brodsky:2009gx}
  S.~J.~Brodsky and R.~F.~Lebed,
  Phys.\ Rev.\ Lett.\  {\bf 102}, 213401 (2009)
  [arXiv:0904.2225 [hep-ph]].
  
  
  \bibitem{Chabysheva:2011ed}
  S.~S.~Chabysheva and J.~R.~Hiller,
  Phys.\ Lett.\  B {\bf 711}, 417 (2012)
  [arXiv:1103.0037 [hep-ph]].
  
  
  \bibitem{deTeramond:2013it} 
  G.~F.~de Teramond, H.~G.~Dosch and S.~J.~Brodsky,
  arXiv:1301.1651 [hep-ph].
  
  
  \bibitem{deTeramond:2012rt}
  G.~F.~de Teramond and S.~J.~Brodsky,
  arXiv:1203.4025 [hep-ph].


\bibitem{Gutsche:2011vb}
  T.~Gutsche, V.~E.~Lyubovitskij, I.~Schmidt, A.~Vega,
Phys.\ Rev.\ D {\bf 85}, 076003 (2012)
[arXiv:1108.0346 [hep-ph]].

\bibitem{deTeramond:2010ge}
  G.~F.~de Teramond and S.~J.~Brodsky,
  AIP Conf.\ Proc.\  {\bf 1296}, 128 (2010)
  [arXiv:1006.2431 [hep-ph]].


\bibitem{Breitenlohner:1982jf}
  P.~Breitenlohner and D.~Z.~Freedman,
  Annals Phys.\  {\bf 144}, 249 (1982).
  
  
  \bibitem{Karch:2006pv}
  A.~Karch, E.~Katz, D.~T.~Son and M.~A.~Stephanov,
  Phys.\ Rev.\  D {\bf 74}, 015005 (2006)
  [arXiv:hep-ph/0602229].
  
  
  \bibitem{Brodsky:2013yy}
  S.~J.~Brodsky,  G.~F.~de Teramond and H. G. Dosch,
  ``Confinement and Chiral Properties in Holographic QCD,''
 {arXiv:1301.XXXX [hep-ph]}.
 
 \bibitem{PDG2012}
 J.~Beringer {\it et al.} [Particle Data Group], 
 Phys. Rev. {\bf D86}, 010001 (2012).

\bibitem{Diehl:2005wq}
  M.~Diehl,
  Nucl.\ Phys.\ Proc.\ Suppl.\  {\bf 161}, 49 (2006)
  [arXiv:hep-ph/0510221].
  
  
  \bibitem{Arriola:2012vk} 
  E.~R.~Arriola, W.~Broniowski and P.~Masjuan,
  arXiv:1210.7153 [hep-ph].
  
  \bibitem{Brodsky:2011pw} 
  S.~J.~Brodsky and G.~F.~de Teramond,
  PoS QCD {\bf -TNT-II}, 008 (2011)
  [arXiv:1112.4212 [hep-th]].
  
   \bibitem{Baldini:1998qn}
  R.~Baldini {\it et al.},
Eur.\ Phys.\ J.\  C {\bf 11}, 709 (1999).

  \bibitem{Tadevosyan:2007yd}
  V.~Tadevosyan {\it et al.},
  Phys.\ Rev.\  C {\bf 75}, 055205 (2007)
 [arXiv:nucl-ex/0607007];
  T.~Horn {\it et al.},
 Phys.\ Rev.\ Lett.\  {\bf 97}, 192001 (2006)
  [arXiv:nucl-ex/0607005].
  
  
  \bibitem{Aubert:2009ad} 
  B.~Aubert {\it et al.} 
  Phys.\ Rev.\ Lett.\  {\bf 103}, 231801 (2009)
  [arXiv:0908.3589 [hep-ex]].
  
  
  \bibitem{Seth:2012nn} 
  K.~K.~Seth, S.~Dobbs,  {\it et al.},
  arXiv:1210.1596 [hep-ex].
  
   \bibitem{Kuhn:1990ad} 
  J.~H.~Kuhn and A.~Santamaria,
 Z.\ Phys.\ C {\bf 48}, 445 (1990).
   
 
 \bibitem{Bruch:2004py} 
  C.~Bruch, A.~Khodjamirian and J.~H.~Kuhn,
Eur.\ Phys.\ J.\ C {\bf 39}, 41 (2005)
  [ hep-ph/0409080].
  
  
  \bibitem{Hanhart:2012wi} 
  C.~Hanhart,
Phys.\ Lett.\ B {\bf 715}, 170 (2012)
  [arXiv:1203.6839 [hep-ph]].
  
   \bibitem{dTB}
  G.~F.~de Teramond and S.~J.~Brodsky (In preparation).


\bibitem{Appelquist:1977tw}
  T.~Appelquist, M.~Dine and I.~J.~Muzinich,
  Phys.\ Lett.\  B {\bf 69}, 231 (1977).
  


\end{thebibliography}
\end{document}